\documentclass[twocolumn,showpacs,amsmath,amssymb,prl]{revtex4-1}  

\usepackage{graphicx}
\usepackage{dcolumn}
\usepackage{bm}
\usepackage{epstopdf}

\begin{document}

\title{Observation of the Presuperfluid Regime in a Two-Dimensional Bose Gas}
\author{S. Tung, G. Lamporesi, D. Lobser, L. Xia, and E.~A. Cornell\footnote{NIST}}
\altaffiliation{Quantum Physics Division, National Institute of Standards and
Technology, Boulder, CO, USA}
\affiliation{ JILA, National Institute of Standards and Technology
and University of Colorado, and Department of Physics, University of
Colorado, Boulder, Colorado 80309-0440, USA}
\date{\today}

\begin{abstract}
In complementary images of coordinate-space and momentum-space
density in a trapped 2D Bose gas, we observe the emergence of
presuperfluid behavior. As phase-space density $\rho$ increases
toward degenerate values, we observe a gradual divergence of the
compressibility $\kappa$ from the value predicted by a bare-atom
model, $\kappa_{ba}$. $\kappa/\kappa_{ba}$ grows to 1.7 before
$\rho$ reaches the value for which we observe the sudden emergence
of a spike at $p=0$ in momentum space. Momentum-space images are
acquired by means of a 2D focusing technique.   Our data represent
the first observation of non-mean-field physics in the
presuperfluid but degenerate 2D Bose gas.
\end{abstract}

\maketitle

Because of the enhanced role of fluctuations in low-dimensional
systems \cite{Mermin}, a two-dimensional (2D) Bose gas at nonzero
temperature does not have long-range phase coherence. In a
homogeneous system there can be at best only a quasicondensate, no
true Bose-Einstein condensation (BEC). Under the combined effect of
interactions and quantum degeneracy, however, there is nonetheless a
phase transition known as Berezinskii-Kosterlitz-Thouless (BKT)
associated with the unbinding of vortex pairs \cite{BKT}. Below the
critical temperature $T_{\mathrm{BKT}}$, the system is superfluid.
\par
Experiments in 2D atomic gases \cite{Hadzibabic,Kruger, Rath,
Phillips} are usually conducted in the presence of an
inhomogeneous trapping potential.  In the complete absence of
interactions, the confining potential can resurrect a traditional
BEC \cite{Bagnato}, but for realistic experimental parameters,
interatomic interactions tend to suppress BEC by smoothing out
the spatial profile \cite{Petrov,
Hadzibabic,Hadzibabic2,Kruger,Phillips, Rath} of the mean density
to the point where the sample can be understood as a collection of
locally uniform spatial regions, each of which is characterized
by a particular local density and thus a particular local value
of $T_{\mathrm{BKT}}$. Although these local regions may be too
small to test in detail the coherence-related predictions of BKT
theory, qualitative effects have been observed in experiment
\cite{Hadzibabic, Phillips}.
\par
Our particular interest is in the region just to the warm side of
$T_{\mathrm{BKT}}$. In an earlier experiment on bosons trapped in a
2D optical lattice, we observed a proliferation of vortices as we
warmed through the discrete-case equivalent of $T_{\mathrm{BKT}}$
\cite{Schweikhard}. But in that experiment a great many mesoscopic
condensates were present, one at each lattice site, on both sides of
the BKT transition, because they had condensed at a
$T_{\mathrm{BEC}}$ distinct from and well above $T_{\mathrm{BKT}}$.
For the continuous case, in contrast, there is no corresponding
second transition temperature above $T_{\mathrm{BKT}}$. But if the
cooling gas has by $T_{\mathrm{BKT}}$ already become a medium that
can support vortices, whether bound or not, then heuristically we
see that it must have continuously evolved from a fully fluctuating
nondegenerate gas into a sort of presuperfluid with suppressed
density fluctuations \cite{Hadzibabic3}. Theory
\cite{Petrov,Holzmann1,Blakie,Prokofev,Holzmann2, Hadzibabic3}
validates this intuition, and experiments \cite{Rath} have in turn
been consistent with predictions of that theory.  Up until now,
however, experiments have not been directly sensitive to the
properties of the presuperfluid, $T \gtrsim T_{\mathrm{BKT}}$ gas.
The goal of the present work is to provide a minimal-assumption,
first empirical look at this exotic regime. We emphasize key
features of this approach: 1. Our line of sight is along the axis of tight confinement: we do
not need to do a deconvolution of our images to get the 2D density
distribution.  Steps are taken to minimize systematic errors in
density measurements. 2. We analyze our \textit{in situ} images to extract
the local compressibility, a quantity directly sensitive to local
microscopic physics. 3. We use a 2D focusing technique to record
high-resolution 2D momentum-space images complementary to the
coordinate-space images. We make corresponding inferences about
nonlocal coherence. 4. We use a simple but robust ``bare-atom" model to
correct the observed density for the presence of a small population
in excited states in the tight confinement direction, and to
determine a bare-atom compressibility with which to compare our
observations.
\par
Experimentally, we create a stack of well-isolated quasi-2D layers
by superimposing a one-dimensional, blue-detuned optical lattice
with lattice spacing 3.8 $\mu m$ onto a magnetically trapped,
evaporatively cooled cloud of Rb-87 atoms. Within each layer,
approximately $6.2\times10^5$ atoms feel a harmonic potential
characterized by frequencies
$(\omega_{r},\omega_{z})$=2$\pi$(10,1400) Hz. The characteristic
dimensionless 2D interaction strength is
$\tilde{g}=\sqrt{8\pi}(a_s/a_{ho})=0.093$,  where $a_s$ is the 3D
scattering length and $a_{ho}$ is the $\hat{z}$ harmonic-oscillator
length \cite{gtilde}. The atoms are allowed to equilibrate in their
2D geometry before probing occurs. Right before probing the
resulting coordinate- or momentum-space distribution, we apply a microwave
pulse together with a transient magnetic field to pump atoms in the central
layer into another hyperfine state, resonant with the probe light.
The strength of the
microwave selection pulse is adjusted to keep the peak optical
density of the imaged fraction within the linear dynamic range of
our imaging.
\par
Our momentum-space imaging makes use of a focusing technique which
is an extension to 2D of a procedure developed for imaging 1D
momentum distributions \cite{Walraven,Druten}. It yields a much
cleaner separation of momentum and coordinate-space distributions
than is obtained in earlier experiments \cite{Phillips, Kruger}.
After selecting a single layer, we turn off the optical lattice and
let the layer expand into a purely magnetic trap with frequencies
$(\omega_{r},\omega_{z})$=2$\pi$(5.2,10.4) Hz. Because of the 140:1
aspect ratio of the cloud, the expansion is initially purely axial,
very rapidly reducing the 3D density while not affecting the
in-plane coordinate- or momentum-space distributions. After this
near-instantaneous suppression of the repulsive atom-atom
interactions, each atom undergoes free harmonic motion in the $x-y$
plane. After a dwell time $t=\frac{1}{4}\frac{2\pi}{\omega_r}$, just
as the initial 2D momentum- and coordinate-space distributions have
swapped, we take an absorption image [Fig.~1(a)]. Scaling the
spatial coordinate by $m\omega_r$ yields the momentum-space
distribution that existed just as the probe sequence began.   We
take azimuthal averages of the absorption images before plotting and
fitting the data.

\par
For coordinate-space images the procedure is similar, but
after the lattice turn-off, we wait only 1 ms before taking
the absorption image [Fig.~1(b)]. The 2D density remains essentially
frozen while the 3D density -- and related imaging artifacts
\cite{Rath} -- are much reduced.

\begin{figure}
   \includegraphics[width=0.45\textwidth]{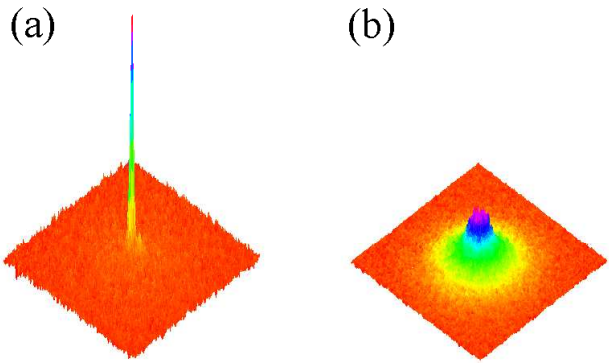}
   \caption{(a) A trap-focused, momentum-space
   image. (b) An in-situ, coordinate-space image. The corresponding azimuthal averages are shown in
    Figs. 2(g) and 2(c). }
   \label{fig1}
\end{figure}

\par

To extract a signal for many-body physics from our data, we compare
our data to a fully fluctuating, bare-atom model, in essence the
Paris group's mean-field, Hartree-Fock, local-density model
\cite{Hadzibabic2}. The mean occupation of a single-atom state $k$
is given by the Bose-Einstein distribution,
$N_{k}=\frac{1}{e^{(E_k-\mu)/k_BT}-1}$, where $E_k$ is the energy of
the state. For our system, $k_BT \gg \hbar\omega_{r}$, but $k_BT
\sim \hbar\omega_{z}$. We treat the atomic motion semiclassically in
the in-plane direction, while preserving discrete
harmonic-oscillator levels in the $\hat{z}$ direction. The 2D
coordinate-space density in the $j$th axial level is
\begin{equation}
n_j(\vec{r})=\frac{1}{h^2}\iint \mathrm{d}^{2}\vec{p}
\frac{1}{e^{[\varepsilon(\vec{p})+\theta_j(\vec{r})-\mu_j(\vec{r})]/k_BT}-1},
\label{eq2}
\end{equation}
where the free particle dispersion
$\varepsilon(\vec{p})=\frac{p^2}{2m}$. The local chemical
potential for the $j$th level is given by
\begin{equation}
\begin{split}
\mu_j(\vec{r})& =\mu^{global}-\frac{1}{2}m\omega_r^2r^2- j \hbar\omega_z \\
&-\sum\limits_{l\neq j}2\left(\frac{4\pi\hbar^2}{m} a_s f_{jl}
n_l(\vec{r})\right), \label{eq3}
\end{split}
\end{equation}
whereas the intralevel interaction energy is
\begin{equation}\label{eq4}
\theta_j(\vec{r})=2(4\pi\hbar^2a_s/m) f_{jj}n_{j}.
\end{equation}
The relevant mean-field interaction energies depend on $f_{jl}$
which are the normalized density overlap integrals over the axial
dimension between densities associated with axial quantum states
$j$ and $l$. Interaction energies are comfortably less than the
axial spacing $\hbar\omega_{z}$, justifying our treating the axial
wave functions as frozen. We define a quantity $u_{00} \equiv
(4\pi\hbar^2a_s/m)f_{00}$, such that we can write
$\theta_0(\vec{r}) = 2 u_{00} n_0 = 2 (\hbar^2\tilde{g}/m)
n_0(\vec{r})$. Evaluating the integral in Eq. \eqref{eq2}, we get
\begin{equation}
n_j(\vec{r})=-\ln(1-e^{-[\theta_j(\vec{r})-\mu_j(\vec{r})]/k_BT})/\lambda_{db}^2
\label{eq5}
\end{equation}
where the de Broglie wavelength $\lambda_{db}
=\sqrt{2\pi\hbar^2/mk_BT}$. For any given value of $\mu^{global}$
and $T$, $n_j$ are determined self-consistently. For $k_BT
\lesssim \hbar\omega_z$, the model converges in just a few
iterations.

\par
The bare-atom model is a no-condensate model from which all the
many-body effects associated with degenerate bosons has been
intentionally omitted: the additional factor of 2 in front of the
parentheses in Eq.~\eqref{eq3} and \eqref{eq4} arises from an
implicit assumption that the second-order correlation function at
zero distance is 2, as it would be for fully fluctuating,
nondegenerate ideal bosons, and not 1, as for a 3D BEC.
Furthermore, $\varepsilon(\vec{p})=p^2/2m$ is the dispersion
relation for independent atoms moving in a mean-field potential.
There are no collective excitations such as phonons.
\par
All the same, the bare-atom model should do very well where
phase-space density $\rho_j \equiv n_j \lambda_{db}^2 < 1$, true for
$j>0$ in our system. As for the calculated value of $n_0(\vec{r})$,
this will begin to fail for $\rho_0 \gtrsim 3.5$, but a comparison
observations with the naive, bare-atom $n_0$ will allow us to
quantify the telltale discrepancy.
\par
\begin{figure}
 \includegraphics[width=0.45\textwidth]{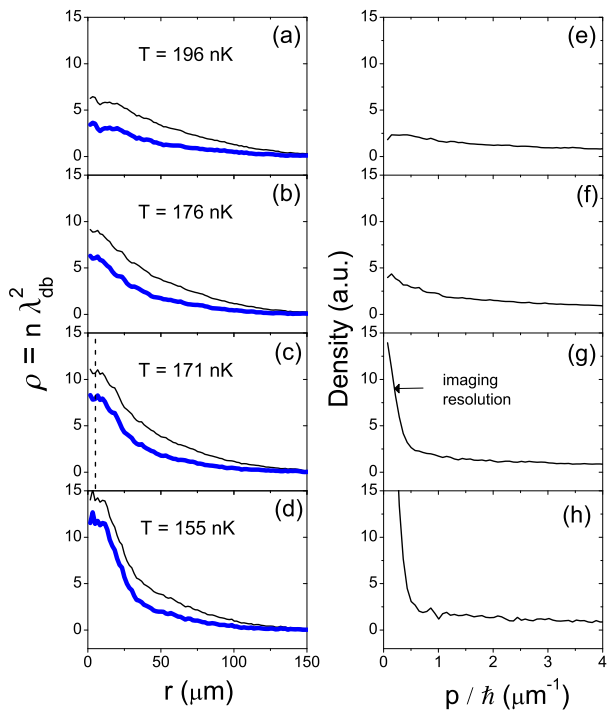}
 \caption{(a)--(d) Coordinate-space distributions and (e)--(h) corresponding
 momentum distributions.  Two distributions in the same row are
 taken under near-identical conditions. The thin black curves give
 the azimuthal averages of $n_{meas}$ from the raw images. The thick
 blue curves in the coordinate-space distributions are the ground-state 
 distribution $n_0$ after correcting for $n_{k>0}$. The spike
 in momentum space that first appears in (g) has no corresponding dramatic
 change in coordinate space (c). The vertical dotted line in (c)
 represents the inverse of the momentum resolution limit indicated in
 (g) and is thus a lower limit on the spatial extent of the coherence
 of the  population of low-p (high-coherence) atoms represented by the area
 (about 1.4 $\%$ of total) under the spike in (g).}
 \label{fig2}
\end{figure}

An analysis of a coordinate-space image proceeds as follows. We measure the 
integrated density in the $z$-direction with contributions from all populated 
excited axial levels. $n_{meas}(\vec{r}) = \sum_{j} n_j (\vec{r})$, but the interesting 
physics lies in $n_0(\vec{r})$. We compare the results of the bare-atom model to 
observed $n_{meas}$ and fit the parameters $T$ and $\mu^{global}$ to the low 
phase-space (hence well-understood) regions of the cloud. With the chemical 
potential and temperature obtained from the fit, we can use Eq. \eqref{eq5} to 
evaluate the excited state distributions. Then we numerically find a self-consistent 
solution to get $n_{k}$ ($k > 0$), with the constraint $n_{0}=n_{meas}- \sum_{k>0} n_k$; see Figs.
\ref{fig2}(a)--\ref{fig2}(d).

\par

\begin{figure}
\includegraphics[width=0.4\textwidth]{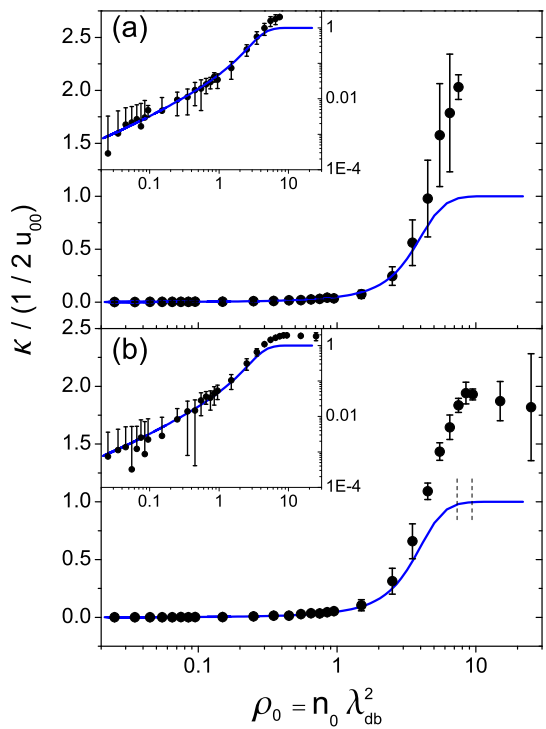}
\caption{Scaled compressibility $\kappa$ vs phase-space density $\rho_0$.
  (a) Measured $\kappa$ extracted from images of samples at $T = 171$ nK,
  as in the image in Fig.2(c). Black circles are data averaged over
  the values calculated from images of three separate clouds.
  The blue curves are $\kappa_{ba}$ calculated from the
  bare-atom model. (b) Same but with $\kappa$ extracted
  from images of samples at $T = 128$ nK. The paired vertical
  dotted lines indicate the location of the jump in coherence
  discussed in the text. \label{fig3}}
\end{figure}

Once the corrected ground-state distribution $n_{0}$ is
extracted from $n_{meas}$,  we calculate the scaled (by $n^2$) isothermal
compressibility $\kappa$ at each imaging pixel
\begin{equation}
\kappa =d n_0/d \mu_{0}=(d n_0/d r)(d \mu_0/d r)^{-1}. \label{eq6}
\end{equation}
Although $\mu^{global}$ is not a quantity we can know with great
accuracy in a model-independent way, $d\mu_0 / d r \approx -
d(\frac{1}{2}m\omega_r^2r^2)/dr = - m \omega_r^2 r$ is known
quite precisely, as the contributions to $\mu_0$ arising from the
mean field of axially excited atoms are small and correctable.
$dn_0/dr$ is determined from our coordinate-space images. Equation 
\eqref{eq6} then gives $\kappa$ at each discrete radius in an
image. We plot the result vs the local phase-space density
$\rho_0=n_0 \lambda_{db}^2$ in Fig. \ref{fig3}. We compare $\kappa$ to the
value $\kappa_{ba}$ that the bare-atom model would predict
at the same density. For an observed value of $n_0$, we
numerically solve the bare-atom prediction $n_0 = -\ln
[1-exp(\mu_0 - \theta_0 (n_0))/k_BT]/\lambda_{db}^2$ for $\mu_0$,
and determine how $n_0$ changes for small changes in $\mu_0$, and
thus extract $\kappa_{ba}$ (Fig. \ref{fig3}).
\par
As a test of the local-density approximation that is central to
our analysis, we determine $\kappa$ using images from two very
different classes of samples: clouds with $T$ = 171 nK and central
$\rho_0$ of about 9  [Fig. \ref{fig3}(a)], and clouds with $T$ =
128 nK and central $\rho_0$ of about 30 [Fig. \ref{fig3}(b)]. The
shape of $\kappa(\rho)$ is the same and, in particular, the value of
$\rho_0$ for which the extracted value of $\kappa$ becomes distinguishable from
$\kappa_{ba}$ is in both cases about 4.
\par
A note on our preferred method of fixing of $\alpha$, the
calibration scale factor that relates $n_{meas}(\vec{r})$ to the
observed optical density profile: we considered (i) an error-prone
calculation based on optical absorption cross section and (ii) a
model-dependent (even unto logical circularity) fitting of
$\alpha$ in the image analysis but settled finally on (iii)
interleaving our data runs with auxiliary measurements of very
low $T$ clouds in which the atoms are in a near-pure Thomas-Fermi
inverted parabola with negligible noncondensed wings.  In this
limit, we assume
$\mu^{global}$=$u_{00}n_0(0)$ and thus fix $\alpha$. This assumption means that
our measured values of $\kappa$, scaled as in Fig.
\ref{fig3}, are constrained to saturate to an average of 2.0 at
very high $\rho_0$. In essence, we get accurate measurements of
the sample density and temperature in the exotic, intermediate
regime of $\rho_0$, by assuming prior good
understanding of behavior in the experimentally well-characterized regimes
of low degeneracy, mean-field at high $T$
and of high-coherence, pure condensates at low $T$.

\par
In Fig. \ref{fig2}, we present side-by-side pairs of
coordinate-space and momentum-space distribution taken in a
sequence such that pairs represent images of clouds with very
similar temperatures and total atom number, such that the preimaging values of $\rho_0(0)$ are the same between pairs to within 10\%.  As $T$ decreases,
there is no obvious sudden change in the coordinate-space
distribution $n_0(\vec{r})$, [Figs. \ref{fig2}(a)--(d)], while in
momentum space [Figs. \ref{fig2}(e)--(h)] a central spike suddenly
emerges at $T$ =171 nK [Fig. \ref{fig2}(g)]. The inverse width of
this central spike is a measure of the spatial extent of the
coherent fraction in the highest-density region of the cloud. Our
momentum-space resolution is such that the presence of a
resolution-limited peak implies that at least some coherence
extends over a central disk of radius 4.5 $\mu$m, or $\gtrsim 10
\lambda_{db}$. From coordinate-space images taken under the same
conditions for which the coherence spike first appears in
momentum space, we determine that it happens when the central
value of $\rho_0$ = 8.0(0.7) (this critical
value $\rho_c$ is determined from looking at many more pairs of
images than are presented in Fig.~\ref{fig2}). We emphasize that
from the coordinate-space distribution alone, the identification
of a transition temperature would require model-dependent
analysis of the smoothly varying distribution, while with access
to both distributions at once, we readily see that a modest
change in the central phase-space density of $<15\%$ causes the
distribution at $p=0$ to jump by a factor of 3.
\par

What have our observations told us about the 2D Bose gas as it
cools towards the BKT transition?  We can say empirically that as
$\rho_0$ varies from about 7.2 to 8.7, we see a dramatic increase
of coherence in a range $>$4.5 $\mu$m, jumping by a factor of
3. The transition may be even sharper, but temporal drifts
limit resolution.  The predicted
\cite{Prokofev} critical value is $\rho_c=ln
(380/\tilde{g})=8.3$.

Our most interesting observation is that in warmer gases, for
$\rho_0 \approx 4$, well before the sudden onset of coherence, we
can already resolve that compressibility is above what a bare-atom
model of degenerate bosons can account for.  As $\rho_0$ reaches its
coherence-jump value, 8.0, $\kappa/\kappa_{ba}$ has already
increased to 1.7 (Fig. \ref{fig3}). It is natural to associate the
increase in $\kappa$/$\kappa_{ba}$ with the gradual changing of the
interaction energy from its value in a fully fluctuating gas, $2
u_{00} n_0$, to its fully condensed value, $u_{00} n_0$, and to draw
a corresponding inference about the zero-range second-order
correlation function.  For $\tilde{g}=0.093$ the corresponding
theoretical prediction (from Ref. \cite{Prokofev} and Sec.~3.3 of
Ref. \cite{Hadzibabic3}) for $\kappa$/$\kappa_{ba}$ at $\rho_c$ would
be 1.59, in reasonable agreement with our observed 1.7. As a caveat,
our observation of an anomalous $\kappa/\kappa_{ba}$ establishes
definitively only the breakdown of the bare-atom model, which could
be partially due to the violation of the other key bare-atom
assumption, that in-plane excitations correspond to individual atoms
with kinetic energy $\epsilon(\vec{p}) = p^2/2m$. In any case, our
data provide a first definitive observation of non-mea-nfield physics
in the  presuperfluid 2D Bose gas.

We are very pleased to acknowledge useful conversations with Z.
Hadzibabic, J. Dalibard, W. Phillips, M. Holzmann, C. Chin, A.
Imambekov, L.-K. Lim and V. Gurarie.  This work was supported by
NSF and ONR.

{\it Note added.}---Recently, a relevant experimental preprint appeared \cite{Hung}.



\end{document}